\documentclass[prl,aps,amssymb,showpacs,twocolumn]{revtex4}
\usepackage{amsmath}
\usepackage{amssymb}
\usepackage{amsthm}
\usepackage{amsfonts}
\usepackage{algorithmic}
\usepackage{enumerate}
\usepackage{latexsym}
\usepackage[dvips]{graphicx}

\newcommand{\beq}{\begin{equation}}
\newcommand{\eneq}{\end{equation}}

\input{epsf}

\begin{document}

\tolerance 10000

%\draft

\title{On Vortices and Phase Coherence in High $T_c$ Superconductors}

\author { Bogdan A. Bernevig$^\dagger$, Zaira Nazario$^\dagger$, and
David I. Santiago$^{\dagger, \star}$ }

\affiliation{ $\dagger$ Department of Physics, Stanford
University,
         Stanford, California 94305 \\ $\star$ Gravity Probe B Relativity
Mission, Stanford, California 94305}
\begin{abstract}
%\vspace*{-1.0truecm}
\begin{center}

\parbox{14cm}{ We show that an array of Josephson coupled Cooper paired planes 
can never have long range phase  coherence at any finite temperature due to an 
infrared divergence of phase fluctuations. The phase correlations decay in a 
slow enough manner providing enough local phase coherence as to make possible 
the nucleation of vortices. The planes then acquire Kosterlitz-Thouless 
topological order with its intrinsic rigidity and concomitant superfluidity. We
thus conclude that the high temperature superconducting cuprates are 
topologically ordered superconductors rather than phase ordered 
superconductors. For low enough superfluid densities, as in the underdoped 
cuprates\cite{uem}, the transition temperature, $T_c$, will be proportional to 
the superfluid density corresponding to vortex-antivortex unbinding, and not to
disappearance of the Cooper pairing amplitude. Above $T_c$, but below the BCS 
pairing temperature $T_p$, we will have a  dephased Cooper pair fluid that is a
vortex-antivortex liquid. The AC and DC conductivities measured in this region 
are those corresponding to flux flow\cite{bozo}. Furthermore there will be 
vortices above $T_c$ which will lead to Nernst vortexlike response and there 
will be a measurable depairing field $H_{c_2}$ above $T_c$ as evidenced by 
recent experiments\cite{ong1,ong2}.}

\end{center}
\end{abstract}

\pacs{74.20.-z, 74.20.Mn, 74.72.-h, 71.10.Fd, 71.10.Pm }

\date{\today}

\maketitle

In the present letter we propose that the high $T_c$ superconducting cuprates 
are {\it topologically ordered} superconductors exactly like the two 
dimensional (2D) Kosterlitz-Thouless\cite{kt} superfluids {\it and not} long 
range phase ordered superconductors like regular 3D bulk superconductors. We 
show that thermal phase fluctuations destroy phase coherence for {\it even} an 
infinite array of Josephson coupled superconducting planes. One could naturally
conclude that such systems cannot be superfluid. This conclusion would be 
erroneous as the planes can acquire the rigidity necessary for supporting 
persistent supercurrents through the topological order originally discovered by
Kosterlitz and Thouless\cite{kt}.

High-temperature superconductors exhibit the pseudogap 
phenomenon\cite{timusk,shen}. In underdoped cuprates, the transition 
temperature, $T_c$, out of the superconducting state is proportional\cite{uem} 
to the superfluid density, $n_s$. This is consistent with the transition being 
an order parameter phase instability\cite{emery} and not a pairing amplitude 
instability as in conventional BCS superconductors. 

The superconducting cuprates are quite anisotropic materials. In the
continuum approximation, a 3D superconductor cannot lose long range phase 
coherence due to thermal phase fluctuations (see Appendix) regardless of 
anisotropy and superfluid density. In 3D, superconductors can lose phase 
coherence if the Cooper pairs have a Bose condensation temperature, $T_c$, 
which is lower than the pairing temperature, $T_p$. This would 
predict\cite{schaf} the relation $T_c \sim n_s^{2/3}$ contrary to 
experiment\cite{uem}.

The superconducting cuprates consist of Copper-Oxygen planes well separated in 
space where all the action happens. They are layered superconductors described
as an array of discrete Josephson coupled planes: a discrete Lawrence-Doniach 
model\cite{vortex,ld}. We obtain that such an array of coupled planes does not 
have long range phase coherence (no ODLRO)\cite{odlro} at any finite 
temperature (see Appendix) due to an infrared divergence of thermal phase 
fluctuations. This could lead one to conclude that superconductivity is 
impossible for such a system as the Meissner effect and infinite conductivity 
are consequences of the phase rigidity of the ordered ground state. 

On the other hand, in 2D there also cannot exist off-diagonal long range order 
(ODLRO) at any finite temperature due to the infrared divergence of
thermal phase fluctuations\cite{rice} (see Appendix). The long distance phase 
correlations of a discrete layered superconductor decay in the same manner as 
for 2D superconductors.  

In 2D quantum states corresponding to different
flows are topologically distinct\cite{kt}. When an energy barrier 
prevents tunneling or thermal hopping between topologically distinct states, 
the system is topologically ordered. This topological order gives rigidity to 
the superfluid ground state and stability to persistent currents. Such a 2D 
superfluid is said to be in the Kosterlitz-Thouless (KT) phase.

The KT phase raises interesting questions about the existence of vortices. We 
now review the old answers to these questions. In superconductors, it is 
impossible to nucleate vortices in the absence of a pairing amplitude. The more
important concern is if one can nucleate vortices in the absence of phase 
coherence. The screening supercurrent intrinsic to a vortex excitation 
precludes the possibility of vortex nucleation in the absence of phase 
coherence at least up to some finite distance. 

In 2D, despite the absence of ODLRO, the phase correlations decay slowly 
enough\cite{rice} to provide sufficient short range phase coherence as to 
make possible the existence of vortices\cite{kt}. In superconductors, a similar
phenomenon occurs in a vortex liquid phase\cite{vortex} in which vortex flow 
destroys ODLRO, yet there is enough short range order to nucleate vortices. The
vortex degrees of freedom and a particular form of their interaction are 
necessary for the stability of the topologically ordered KT phase\cite{kt}.

The KT phase is a vortex insulating fluid of vortex-antivortex bound 
``dipoles''. The inverse distance attraction, i.e. logarithmic interaction, 
among vortices and antivortices is responsible for dipole binding and thus for 
an energy barrier to excite free vortices in the KT phase. The scarcity of
free vortices at low enough temperature provides for the stability against 
decay of supercurrents and ultimately for the topological ordered ground state 
and its concomitant rigidity. The KT phase becomes thermodynamically unstable 
when entropy gain makes vortex-antivortex unbinding favorable at a temperature,
$T_c$, which is proportional to $n_s$, the superfluid 
density\cite{kt,mac,doniach,halp}.

At first, the KT phase was not believed to exist for 2D superconductors as 
vortices interact through an inverse square force law when sufficiently far 
apart\cite{pearl}. Soon after, it was remembered that vortices within a 
penetration depth of each other interact logarithmically with 
distance\cite{mac}. Therefore the KT phase can occur\cite{mac} 
for 2D superconducting films with large penetration depths. For
superconductors with small enough superfluid density, $T_c$ is smaller than 
the BCS Cooper pairing temperature, $T_p$\cite{mac,doniach,halp}. Above $T_c$ 
and below $T_p$ there is a dephased Cooper pair fluid which {\it is not} 
superconducting: it is not a perfect diamagnet nor does it exhibit 
resistanceless conduction. The system thus exhibits the pseudogap phenomenon.

The dephased superfluid, existing above $T_c$ and below $T_p$, is actually
a neutral vortex-antivortex ``metal'' or ``plasma''. The resistivity for this 
dephased superfluid is provided by flux flow in the vortex-antivortex fluid
present at those temperatures\cite{doniach,halp}. Moreover, a magnetic field
acts as a chemical potential creating an imbalance between $+$ and $-$ vortices
so that the vortex-antivortex plasma is no longer neutral\cite{doniach,minn}. 
We have a vortex liquid in the presence of a magnetic field. In the presence of
a thermal gradient these vortices will flow with their Josephson electric field
producing a Nernst effect signal. It also follows that there is a depairing 
field $H_{c_2}$ {\it above} $T_c$\cite{doniach} which collapses to zero at 
$T_p$.

In underdoped BSCCO, measurements of the DC conductivity above $T_c$ agree with
the flux flow conductivity ($k_B T/n_f D \Phi_0^2$ with $\Phi_0 = hc/2e$ the 
flux quantum and $D$ the vortex diffusivity)\cite{bozo} expected of a vortex 
liquid. The behavior just described is {\it exactly} that of the phase above 
$T_c$ and below the pairing temperature, $T_p$, in a KT 
superconductor\cite{kt,mac,doniach,halp,minn}. Furthermore, the AC conductivity
has the universal scaling form expected above a KT superconducting 
phase\cite{bozo}.

Recently there have been reports of vortexlike Nernst effect signals
characteristic of a vortex liquid phase in underdoped high temperature
superconducting cuprates\cite{ong1,ong2} {\it well above} T$_c$.  In 
particular, N. P. Ong and collaborators\cite{ong2} also measure $H_{c_2}$ for 
underdoped cuprates {\it above} T$_c$.  From $H_{c_2}$ the coherence length was
determined according to $\xi_0 = \sqrt{\Phi_0 / 2 \pi H_{c_2}}$. From 
photoemission experiments\cite{ding} the coherence length was determined 
according to $\xi_p \simeq \hbar v_F /\Delta_0$, with $\Delta_0$ the  maximum 
gap. The two lengths agree closely with each other and track each other with
doping. Mobile vortices and a depairing field above $T_c$ is the behavior of a 
dephased KT superfluid below the pairing temperature $T_p$\cite{doniach}.

In summary, we have shown that layered superconductors like the high 
temperature superconducting planes cannot have long range phase coherence
at any finite temperature just like 2D superfluids. For low enough superfluid 
densities, as in the underdoped regime, the transition out of the 
superconducting state is through destruction of topological order by 
vortex-anitvortex unbinding at a temperature $T_c \sim n_s$ smaller than the 
Cooper pairing temperature $T_p$, thus exhibiting the pseudogap phenomenon
just like 2D KT superconductors. Above $T_c$ and below $T_p$, there is 
dephased Cooper pair fluid whose dynamical response is mainly that of a 
vortex liquid in the same manner as KT superconductors.

We thus conclude that layered superconductors like the superconducting cuprates
are topologically ordered Kosterlitz-Thouless superconductors. Just like in 2D,
the presence of vortex-antivortex dipoles at low enough temperatures stabilizes
Kosterlitz-Thouless topological order which provides the necessary rigidity for
a superconducting ground state. In the absence of this topological order, 
superconductivity would be {\it impossible} for 2D and for layered 
superconductors due to the absence of ODLRO. 

A layered KT superconductor above its KT transition temperature will consist
of planes of a dephased Cooper pair fluid with finite resistance for transport
in the plane. For interplane charge transport, electrons have to be ripped
out of the Cooper pair fluid and hopped from plane to plane. Thus c-axis 
transport will  exhibit a ``semiconducting'' gap. This behavior is the one
observed in the superconducting cuprates in AC conductivity 
measurements\cite{ir}.

The physics proposed in this work should be universal for all dephased KT 
superfluids. In particular, artificially engineered dephased superconducting 
systems as studied in the experiments of Dynes and collaborators\cite{dynes} 
should exhibit similar properties. In such experiments we thus predict the 
existence of vortex excitations {\it above} $T_c$ on the ``underdoped'' regime 
in the presence of a magnetic field. If the granularity does not cause strong 
enough pinning, these could be observed via Nernst signals as we then expect 
the vortex phase to be a vortex liquid characteristic of a dephased KT 
superfluid as in the Ong experiments\cite{ong1,ong2}. If the vortices are 
pinned and hence not mobile, they should be observable with an STM. We also 
predict that a depairing field $H_{c_2}$ can be measured above $T_c$. 

{\bf Acknowledgments} We would like to thank M. Beasley, T. Cuk, B. Gardner,
and A. Silbergleit for interesting and helpful discussions. Bogdan A. Bernevig 
was supported through the Stanford SGF program.  Zaira Nazario was supported by
The School of Humanities and Sciences at Stanford University. 
David I. Santiago was supported by NASA grant NAS 8-39225 to Gravity Probe B.

\appendix

\section{Appendix: Off-Diagonal Long Range Order}

In this Appendix we study the effect of thermal phase fluctuations on long
range phase coherence or ODLRO\cite{odlro} in anisotropic 3D superconductors, 
2D superconductors, and layered superconductors. We will work at the level of 
Ginzburg-Landau\cite{gl} (GL) concentrating in a Ginzburg-Landau 
{\it phase only} model. Thus the GL order parameter is taken to be $\psi
(\vec x)=\psi_0 e^{i \phi(\vec x)}$ where $\psi(\vec x)=\psi_0 =\sqrt{n_s}$ is 
the order parameter amplitude that minimizes the GL free energy.

The GL phase only Hamiltonian is given by

\begin{equation}
\text{$\cal{H}$}[\varphi(\vec x)]=
\frac{\hbar^2 \psi_0^2}{2m}\int [(\nabla_{\perp}\varphi(\vec x))^2
+\frac{m}{M}(\partial_z\varphi(\vec x))^2]\, d\vec x 
\end{equation}

\noindent where $\nabla_{\perp}$ is the gradient operator in the x-y plane.
Since the cuprates are anisotropic, in order to model them by a continuous GL 
model, we introduce anisotropy by considering a different mass in the z 
direction than in the x-y direction\cite{ld}. The separation between the planes
$c$ is larger than the Copper-Oxygen plane lattice constant $a$.  

We test for the existence of ODLRO by studying the fixed point properties of
the order parameter correlation function $G(\vec x_1- \vec x_2) = 
<\psi(\vec x_1)\psi^*(\vec x_2)>$. There will be ODLRO if the correlation 
function is finite as $|\vec x_1 - \vec x_2|\rightarrow \infty$ 
due to broken gauge invariance in the superfluid state\cite{phil}. 

Generalizing to the 3D anisotropic superconductor, and stealing from the 
results of Rice and others\cite{rice}, we have

\begin{equation*}
G(\vec x_1- \vec x_2) = \frac{1}{Z}\int \prod d\varphi(\vec x)
\psi_0^2
\end{equation*}

\begin{equation}
\times\exp\left(i(\varphi(\vec x_1)-\varphi(\vec x_2))-\beta
\text{$\cal{H}$}[\varphi(\vec x)]\right) \label{gor}
\end{equation}

\noindent where $Z$ is the GL partition function. We define the Fourier 
transform $\varphi(\vec x)=\sum_{\vec k}\Phi_{\vec k}e^{i\vec k\cdot\vec x}$,
with $\Phi_{\vec k} \equiv \varphi_{\vec k}+i\chi_{\vec k}$,
where $\varphi_{\vec k}$ and $\chi_{\vec k}$ are both real. The reality of
$\varphi(\vec x)$ implies $\Phi_{\vec k}=\Phi_{-\vec k}^*$, or
$\varphi_{\vec k}=\varphi_{-\vec k}$  and $\chi_{\vec k}=-\chi_{-\vec k}$.
The correlation function (\ref{gor}), after plenty of massaging, 
becomes\cite{rice}

\begin{equation}
G = \psi_0^2 \exp\left(\frac{-1}{2 V \psi_0^2 \beta
(\hbar^2/2m)}\sum_{\vec k}^{\vec k<Q} \frac{1-\cos(\vec k\cdot\vec
X)}{(k_x^2+k_y^2+(m/M)k_z^2)}\right) \label{cor}
\end{equation}

\noindent where $V$ is the volume of the system, and $\vec X \equiv 
\vec x_1-\vec x_2$. Since fixed point properties are cut-off independent, the 
momentum sums are cut-off at the ultraviolet anisotropically: $Q_{z} = 
\sqrt{M/m}Q_{x \, y} $.

For convenience we make the integrand isotropic by defining $\vec q \equiv 
(k_x, k_y, k_z \sqrt{m/M})$, and $\vec R \equiv (x, y,z \sqrt{M/m})$ so that 
$\vec q\cdot\vec R=\vec k\cdot\vec X$. This allows us to write the sum in the 
exponent as $I=\sum_{\vec q}^{\vec q<Q}\frac{1-\cos(\vec q\cdot\vec R)}{q^2}$. 
To examine the fixed point properties ($Q \rightarrow \infty$), we convert the 
sum to an integral in the thermodynamic limit thus getting $ I=\frac{V}{(2 
\pi)^3} \int^Q_0 \frac{1-\cos(qR\cos\theta)}{q^2}q^2\sin\theta\, d\theta \,
d\phi \, dq $, or $I=\frac{V}{(2 \pi)^3} \int_{0}^{Q} [1-J_0(qX)] d q$,
which as $Q\rightarrow \infty$ yields $I= Q \left(\frac{V }{2 \pi^2}
-\frac{V}{4\pi QR}\right)$. The $Q$ in front of the integral can be subsumed
in an unobservable superfluid density renormalization. The large $Q$ fixed 
point limit is now a finite cut-off independent number for arbitrary $R$. 
Hence the order parameter correlator is finite for infinite separation. 
Therefore there is ODLRO in 3D irrespective of anisotropy and superfluid 
density as long as the superconductor is paired. The transition out of such a 
superconducting state is {\it necessarily} a depairing BCS transition, unless 
the superfluid density is small enough for the Cooper pairs to Bose uncondense 
at a temperature below $T_p$.

In the case of a two dimensional superconductor\cite{rice}, the correlation 
function is

\begin{equation}
G = \psi_0^2 
 \exp\left(\frac{-1}{2 A \psi_0^2 \beta (\hbar^2/2m)}\sum_{\vec
k}^{\vec k<Q} \frac{1-\cos(\vec k\cdot\vec
X)}{(k_x^2+k_y^2)}\right) \label{cor2}
\end{equation}

\noindent where now $A$ is the area of the system. In the thermodynamic limit 
and fixed point limit, $Q>>1$, the sum in the exponent becomes $I= \frac{A}{
(2\pi)^2} \int^Q_0 \frac{1-\cos(kX\cos\theta)}{k^2}k\, dk\, d\theta $, or with 
$\frac{A}{2\pi} \int_{0}^{1}\frac{1-J_0(QXt)}{t} d t$, with $t=q/Q$. The
integral becomes cutoff independent in the limit $QX  \rightarrow \infty$ 
since $J_0(QXt) \rightarrow 0$. In this limit the integral has an infrared 
logarithmic divergence. That is, phase correlations are destroyed by infrared 
thermal phase fluctuations. Therefore there is no ODLRO in the usual sense in 
2D.

In order to examine ODLRO in the cuprates, which are layered superconductors, 
we evaluate the correlator for discrete Josephson coupled planes. We consider 
planes, numbered by $n$, which are separated by a distance $c$ in the $z$ 
direction, so that the vector $\vec{X} = (x,y,z) \equiv (x,y,nc) \equiv 
(\vec{x}_\bot, nc)$. The GL Hamiltonian is given by\cite{vortex,ld}:

\begin{equation*}
\text{$\cal{H}$}[\varphi(\vec x_\bot, nc)]= \frac{\hbar^2}{2m}\int
(\sum_{n=-N_z}^{N_z} ([\partial_{x_\bot}\varphi(\vec x_\bot, nc)]^2 
\end{equation*}

\begin{equation}
+[\frac{\varphi(\vec x_\bot, (n+1)c) - \varphi(\vec x_\bot, nc)}{c}]^2
)\, d\vec x_\bot
\end{equation}

The Fourier transform of the phase is again $\phi(\vec x_\bot, nc)=
\sum_{\vec k}\Phi_{\vec k}e^{i\vec k\cdot\vec x}$
\noindent but this time $k_x = \frac{2 \pi m }{2l_x}, \; m \in [-N_x, N_x]$; 
$k_y =\frac{2 \pi l }{2l_y}, \; l \in [-N_y, N_y]$; $k_z = \frac{2\pi p }
{2l_z}, \; p \in [-N_z, N_z]$; $l_x = N_x a$, $l_y =N_y a$, $l_z = N_z c$ are 
half the length of the sample in the $x$, $y$ and $z$ directions, and $\vec k
\vec X = k_x x + k_y y + k_z nc = \vec k_\bot \vec x_\bot + k_z
nc$.

The correlator for the layered superconductor is

\begin{equation*}
G(\vec x_1 - \vec x_2) =  \psi_0^2 \times
\end{equation*}

\begin{equation}
\exp\left(\frac{-1}{2\beta
(\hbar^2/2m) \psi_0^2V}\sum_{\vec k_\bot, k_z} \frac{1-\cos(\vec
k\cdot\vec X)}{(k_x^2+k_y^2+\frac{2}{c^2} [1-\cos(k_z c)])}\right)
\end{equation}

\noindent  We look for the in-plane correlation function ($\vec X = (\vec
x_{1 \bot} - \vec x_{2 \bot}, 0)$) which are the more relevant ones as they 
are stronger than the out-of-plane ones. The sum over $k_z$ becomes $
\sum_{ k_z} \frac{1}{(k_x^2+k_y^2+\frac{2}{c^2} (1-\cos(k_z c)))}$ or
$\frac{c^2}{2} \sum_{r=-1}^{1}\frac{1}{y^2 + (1-\cos(\pi r))}$ where $r=p/N_z$,
and $y^2 = \frac{c^2}{2}(k_x^2 + k_y^2)$, and the summation is done is steps of
$1/N_z$. In the limit $N_z \rightarrow \infty$ the sum is transformed into an 
integral to yield $\frac{c^2}{2} N_z \int_{-1}^{1} \frac{1}{y^2 + (1-
\cos(\pi r))} dr = l_z c \frac{2}{y\sqrt{2+y^2}}$.

The correlation function now becomes

\begin{equation*}
G(\vec x_1 - \vec x_2) =  \psi_0^2 \times
\end{equation*}

\begin{equation}
\exp\left(\frac{-\sqrt 2 l_z}{2\beta
(\hbar^2/2m) \psi_0^2V}  \sum_{\vec k_\bot}
\frac{1-\cos(\vec k_\bot\cdot\vec X_\bot)}{ \sqrt{k_x^2 +
k_y^2}\sqrt{2+\frac{c^2}{2}(k_x^2+k_y^2)}}\right)
\end{equation}

\noindent Let $q=\sqrt{k_x^2+k_y^2}$. In the thermodynamic limit the sum is 
evaluated by transforming it into an integral with cutoff $Q$. This yields

\begin{equation*}
G(\vec x_1- \vec x_2) = \psi_0^2 \times
\end{equation*}

\begin{equation*}
\exp\left(\frac{-2  \sqrt 2 l_x l_y l_z }{(2\pi)^2 \beta
(\hbar^2/2m) \psi_0^2V} \int_{0}^{Q}
\int_{0}^{2\pi}\frac{1-\cos( q X \cos \theta)}{
q\sqrt{2+\frac{c^2}{2}q^2}} q d q d\theta \right)
\end{equation*}

\begin{equation}
= \psi_0^2 \exp\left(\frac{-1}{8 \pi \beta
(\hbar^2/2m) \psi_0^2} \int_{0}^{Q} \frac{1-J_0(qX)}{
\sqrt{1+(\frac{cq}{2})^2}} d q \right)
\end{equation}

\noindent  In order to study in a controlled manner the fixed point behavior of
the integral in the exponent, we make the change of variables $t=q/Q$ obtaining
$Q \int_{0}^{1} [1-J_0(QtX)]/[\sqrt{1+(\frac{Qct}{2})^2}] d t $. Taking the 
fixed point ($Q>>1$) limit yields $\frac{1}{c}\int_{0}^{1}\frac{1-J_0(QtX)}{t}
d t$. This is exactly the 2D integral for an isolated plane which has a 
logarithmic infrared divergence and becomes cut-off independent for $QX >> 1$. 
Thus the planes decouple, impeding the existence of long range phase coherence 
for layered superconductors.

\end{document}